\documentclass[fleqn,twocolumn]{openjournal} 

\usepackage{mathptmx}
\usepackage[T1]{fontenc}
\usepackage{ae,aecompl}
\usepackage{graphicx}	% Including figures
\usepackage{amsmath}	% Advanced maths commands
\usepackage{amssymb}	% Extra maths symbols
\usepackage{latexsym}
\usepackage{longtable} 
\usepackage[figuresleft]{rotating}
\usepackage{epsf}
\usepackage{hyperref}
\hypersetup{colorlinks=true,
	urlcolor=blue,  % color of external links
	linkcolor=red,  % color of toc, list of figs etc.
	citecolor=blue, % color of links to bibliography
    menucolor=blue, % color for Acrobat menu buttons
    urlcolor=blue}  % color for \url{...} links

\begin{document}

%\title[The Milliquas Catalogue]{\large{The Million Quasars (Milliquas) Catalogue, v8}} 
\title{\large{\textbf{The Million Quasars (Milliquas) Catalogue, \symbol{118}8}}\vspace{15pt}} 
\author{Eric Wim Flesch}
\affiliation{PO Box 15, Dannevirke 4942, New Zealand; eric@flesch.org} 
\email{eric@flesch.org}

\begin{abstract}
Announcing the final release, v8, of the Milliquas (Million Quasars) quasar catalogue which presents all published quasars to 30 June 2023, including quasars from the first releases of the Dark Energy Spectroscopic Instrument (DESI) and the SDSS-DR18 Black Hole Mapper.  Its totals are 907\,144 type-I QSOs/AGN and 66\,026 high-confidence ($\approx$99\% likelihood) radio/X-ray associated quasar candidates.  Type-II and Bl Lac type objects are also included, bringing the total count to 1\,021\,800.  \textit{Gaia}-EDR3 astrometry is given for most objects.  The catalogue is available on multiple sites.
\end{abstract}

\keywords{catalogs --- quasars: general}

%%%%%%%%%%%%%%%%% BODY OF PAPER %%%%%%%%%%%%%%%%%%

\section{Introduction}
The Million Quasars (Milliquas) catalogue has been available on-line since its inception in 2009, and is the only quasar catalogue to have kept abreast of all the latest quasar discoveries from published papers large \& small since the final edition of the V\'eron-Cetty \& V\'eron quasar catalogue \citep[VCV:][13$^{th}$ edition]{VCV}.  With this version v8 of quasars to 30 June 2023, Milliquas now also presents its final edition.  

The criteria for including published quasars, pipeline quasars, and candidates are as given in the Half Million Quasars catalog \citep[HMQ:][]{HMQ} and references therein, but it has evolved somewhat over time and an overview is given below.  The counts of this final edition are 907\,144 type-I quasars and AGN, 45\,816 type-II objects, 2814 blazars, and 66\,026 high-likelihood (pQSO>=99\%) radio/X-ray associated quasar candidates, which totals to 1\,021\,800 objects presented.     

Milliquas can be downloaded from CDS\footnote{https://cdsarc.cds.unistra.fr/viz-bin/cat/VII/294} or from its home page\footnote{https://quasars.org/milliquas.htm} which also provides a FITS file.  Also CDS provides a query page, as does  
NASA HEASARC\footnote{https://heasarc.gsfc.nasa.gov/W3Browse/all/milliquas.html}.  The ReadMe provided at those sites gives essential information about the data.  Figure 1 shows the Milliquas sky coverage.

The sections below describe salient aspects of this catalogue, rules of inclusion and exclusion, data fixes and newly discovered quasars, and how radio/X-ray associations enable the selection of high-confidence photometric quasar candidates.           
     
\begin{figure*} 
\includegraphics[scale=0.55, angle=0]{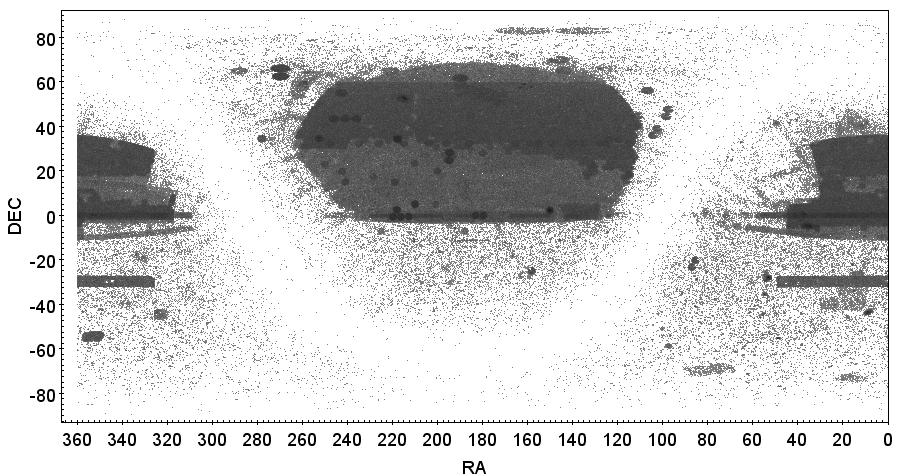} 
\caption{Sky coverage of Milliquas, darker is denser.  The SDSS footprints dominate the North with extensions radiating out from them, and the 2QZ stripe is seen at $\delta=-30^{\circ}$.  DESI-EDR coverage is shown by the high-density beads of coverage sprinkled throughout, including a pair north of +80$^{\circ}$.  The Magellanic Clouds show up at lower right because of multiple surveys to detect quasars behind them.  The large bead at lower left shows the XXL-South AAOmega field.  \textsl{(chart produced with TOPCAT \citep{TOPCAT})} \\ }  
\end{figure*}

\section{Milliquas rationale and selection}  

My original project (\textit{circa} 1999) was to align and overlay radio \& X-ray surveys on to the optical sky.  Quasars of course are prominent in such a task.  I quickly became aware that quasars were often not quite where their authors had placed them, and could be many arcminutes or even degrees offset in extreme cases; my later paper \cite{VCVMQ} gave 380 corrections of 8+ arcsec to the final VCV quasar positions, those fixes subsequently included in the HMQ.  Since the HMQ there have been more such astrometric corrections found, and Table 1 gives 27 new such corrections of 8+ arcsec to legacy quasar positions.  There are 290 more new corrections of 2+ arcsec in the Milliquas data since the HMQ, all individually vetted.  Total presented astrometry is 60.71\% from \textit{Gaia}-EDR3 \citep{GEDR3}, 8.34\% from Pan-STARRS \citep{PS}, 27.79\% from SDSS-Sweeps\footnote{at https://data.sdss.org/sas/dr9/boss/sweeps/dr9/}, 1.69\% from DESI \citep{DES}, 0.37\% from individual authors, and the remaining 1.10\% from the digitized 20$^{th}$ century photographic surveys collected in the ASP optical catalogue \citep{ASP}.  All this shows the strong attention given to astrometric accuracy in the Million Quasars catalogue, both in correctly siting each quasar including all legacy quasars, and presenting those locations with the greatest precision available.          

\begin{sidewaystable}
\scriptsize	 
\caption{Legacy quasars: 27 more positional corrections of 8+ arcsec since the HMQ, in order of \# arcsec moved}
%\fontsize{6pt}{7pt}\selectfont
\tiny
\begin{tabular}{@{\hspace{0pt}}r@{\hspace{4pt}}l@{\hspace{2pt}}c@{\hspace{2pt}}c@{\hspace{4pt}}      
                               c@{\hspace{0pt}}r@{\hspace{6pt}}c@{\hspace{6pt}}l}
\hline 
\# & Name & z & OA ref$^{*}$ & J2000 & moved(") & previous J2000 & comment on move \\
\hline
1 & NGC 4395 U3 & 1.687 & 0081 & 12 19 17.80 +33 29 58.2 & 4897.2 & 12 25 49.33 +33 30 45.0 & OA$^{\dagger}$ described position via offsets without giving co-ordinates, SDSS-DR16Q confirms redshift there, has radio \\
  & & & & & & & ILT J121917.80+332958.3.  VCV used a finding chart from another author which gave a different object with that same name. \\
2 & CXOPS J04219+3300 & 1.125 & 1556 & 04 21 55.03 +33 00 36.1 & 629.4 & 04 21 05.00 +33 00 36.0 & VCV transcription error, OA gave correct position, has X-ray CXO J042155.0+330035. \\
3 & VFQ A4/22 & 1.045 & 1634 & 11 02 05.85 +29 59 14.6 & 212.1 & 11 02 07.20 +30 02 46.0 & apparent OA transcription error, photometry \& redshift match to NBCKv3 candidate 210" south, radio ILTJ110206.01+295914.5 \\
4 & Q 0856+406 & 2.276 & 1860 & 08 59 56.33 +40 24 37.2 & 128.8 & 08 59 45.31 +40 25 04.1 & OA gave approximate position and the spectrum, SDSS-DR16Q spectrum here shows exactly-matching spectral features. \\
5 & NGC 520.D1 & 1.493 & 0088 & 01 25 29.76 +04 11 26.2 & 118.9 & 01 25 25.00 +04 09 51.0 & OA described position via offsets without giving co-ordinates, SDSS-DR16Q confirms redshift there. \\
6 & NGC 615 UB1 & 1.640 & 0081 & 01 34 50.78 -07 21 43.8 & 74.2 & 01 34 52.15 -07 22 55.1 & OA described position via offsets without giving co-ordinates, DESI to get new spectrum there as targetid 39627610338100732. \\
7 & NGC 2916 UB2 & 0.796 & 0081 & 09 35 13.08 +21 47 32.7 & 70.3 & 09 35 13.93 +21 48 42.0 & OA described position via offsets, and SDSS-DR16Q confirms redshift there, but OA finding chart pointed to wrong object. \\  
8 & Q 1409+732 & 3.560 & 0051 & 14 09 48.01 +72 59 29.1 & 67.3 & 14 10 03.20 +72 59 39.0 & OA finding chart (which introduced the article) pointed only to a bad pixel, but OA-given co-ordinates were correct. \\
9 & PKS 1635+159 & 2.145 & 1723 & 16 37 49.04 +15 49 57.4 & 43.7 & 16 37 50.00 +15 49 16.0 & positional correction averred in 1977-ApJ-215-427 without giving co-ordinates, now confirmed by SDSS-DR16Q redshift. \\
10 & PKS 0159+034 & 0.976 & 0670 & 02 01 51.50 +03 43 09.2 & 43.0 & 02 01 50.13 +03 43 47.0 & OA reported quasar as separate from galaxy but VCV kept them conflated for some reason, has radio FIRST J020151.4+034309 \\
11 & RXS J10478-0113 & 0.435 & 0063 & 10 47 49.99 -01 12 43.5 & 41.0 & 10 47 51.66 -01 13 16.0 & OA-given trail from X-ray detection to quasar was not followed by cataloguers, now fixed. \\
12 & TOL 1038.2-27.1 & 1.937 & 0187 & 10 40 34.28 -27 22 30.0 & 40.6 & 10 40 33.20 -27 23 08.0 & OA-given position pointed to nothing, but OA finding chart identified the quasar. \\
13 & KP 1127.9+07.4 & 1.700 & 1756 & 11 30 31.37 +07 12 13.3 & 35.6 & 11 30 32.29 +07 12 46.2 & OA-given position coarse, nearest eligible (prev) object too bright and \textit{Gaia} PM, next (this) one has right magnitude and no PM. \\
14 & ISO J1324-2016 & 1.500 & 1462 & 13 24 47.25 -20 16 12.0 & 21.3 & 13 24 45.73 -20 16 12.0 & OA-given position pointed to nothing, but OA finding chart identified the quasar, has radio RACS J132447.2-201610. \\
15 & Q 0016-357 & 3.199 & 0087 & 00 18 41.60 -35 29 05.4 & 19.7 & 00 18 40.00 -35 29 04.0 & OA-given position \& finding chart were coarse, now radio VLASS J001841.58-352904.9 clarifies position of quasar. \\
16 & Q 13034+2942 & 1.724 & 0222 & 13 05 47.37 +29 26 44.0 & 17.5 & 13 05 46.33 +29 26 33.0 & OA-given position pointed to nothing, but OA finding chart identified the quasar. \\
17 & KP 1229.0+07.8 & 1.930 & 1756 & 12 31 34.53 +07 34 25.8 & 17.1 & 12 31 34.00 +07 34 41.0 & OA-given position pointed to nothing, OA finding chart was coarse but adequate to identify the quasar. \\
18 & LMA 15 & 2.700 & 1080 & 01 05 14.20 +02 11 29.2 & 14.0 & 01 05 13.26 +02 11 29.0 & OA co-ordinates were only to whole time-seconds (i.e., 15 arcsec range), but Pan-STARRS image shows the obvious object. \\
19 & PC 0027+0513 & 3.005 & 1664 & 00 30 26.48 +05 29 53.7 & 12.7 & 00 30 27.06 +05 30 03.0 & OA-given position was near random faint object, but OA finding chart identified the quasar. \\
20 & CT 187 & 1.141 & 0315 & 00 54 13.46 -29 55 44.8 & 11.5 & 00 54 13.66 -29 55 56.0 & OA identified quasar correctly but VCV took co-ordinates of nearby galaxy for some reason. \\
21 & Q 03022-0023 & 2.140 & 0910 & 03 04 46.11 -00 11 27.5 & 10.8 & 03 04 45.93 -00 11 38.0 & Serendipitous but unwanted quasar, OA hurried the J2000, good candidate nearby, X-ray 4XMM J030446.0-001127 confirms. \\
22 & PC 0027+0525 & 4.099 & 1662 & 00 29 49.99 +05 42 14.5 & 10.6 & 00 29 50.00 +05 42 04.0 & OA-given position pointed to nothing, but OA finding chart identified the quasar. \\
23 & CTS A11.17 & 2.800 & 1230 & 22 42 30.72 -36 44 14.4 & 10.5 & 22 42 30.80 -36 44 04.0 & OA warned true positions are often 10" south of given, has radio VLASS J224230.71-364414.0 \\
24 & CTS R05.17 & 2.120 & 1230 & 10 45 36.03 -15 28 41.5 & 9.7 & 10 45 35.60 -15 28 49.0 & OA-given position was near blue doublet of which farther end was picked but has \textit{Gaia}-EDR3 proper motion, near end does not. \\
25 & RX J23360+3023 & 2.094 & 2156 & 23 36 04.08 +30 24 00.2 & 9.2 & 23 36 04.66 +30 23 55.0 & OA confused objects, their pick has proper motion (\textit{Gaia}-EDR3), other does not and has radio VLASS J233604.06+302400.3 \\
26 & S4 1030+39 & 1.095 & 1023 & 10 33 22.03 +39 35 51.0 & 8.8 & 10 33 22.66 +39 35 56.0 & OA conflated bright object with faint true source of X-ray 2SXPS J103322.0+393547 \& radio VLASS J103322.07+393551.0 \\
27 & Q J0641-5050 & 0.618 & 1862 & 06 41 46.00 -50 50 21.7 & 8.2 & 06 41 46.00 -50 50 30.0 & OA identified quasar correctly but VCV somehow took co-ordinates of galaxy to south.  Has X-ray 4XMM J064146.0-505022. \\
\hline
\multicolumn{8}{l}{$^{*}$ references are indexed in the accompanying file "milliquas-references.txt".} \\
\multicolumn{8}{l}{$^{\dagger}$ OA = original author(s) of discovery paper.} \\
\end{tabular}
\end{sidewaystable}

My motivation in all these fixes of legacy positions was not only to present the correct astrometry, but also to preserve the work of the legacy discoverers whose quasars, it seemed, were being submerged by subsequent large surveys which re-surveyed and re-published those quasars without reference to the original discoveries.  The last large survey to carefully identify quasars earlier discovered, was SDSS-DR7Q \citep{DR7Q}, and none since.  Milliquas gives the original discovery citation for all quasars, along with their original names, although, it must be said, the legacy names are mostly as inherited from the VCV catalogue which often changed names according to the quasar co-ordinates.  So within that constraint, I have done my best to preserve the work of the original legacy authors.  

An essential issue is to decide which quasar discoveries should be included in Milliquas, and which not.  The predecessor catalogue VCV \citep{VCV} included questionable objects (so designated by their discoverers) and provided a table of "rejected quasars" which had been included as quasars in earlier editions but had since been found to be not quasars.  The Milliquas approach is different, I include only quasars confidently presented by their discoverers, and objects subsequently found to be not quasars were simply dropped from Milliquas.  There is however a residue of questionable legacy objects held over from VCV data not reviewed.  Workload prohibited a databasing of possible-quasars and non-quasars in the accruing data.  The general rule for this final edition is that all Milliquas v8 objects are 99\% likely to be true quasars (or type-II, or blazars as classified), including the 66\,026 candidates which are calculated in bulk to be 99\% likely to be true quasars -- the candidates selection method is as detailed in HMQ \citep{HMQ} Section 8.       

Some discovery authors present new quasars with a quality flag: one common such flag is 4=confident (2+ strong lines), 3=good (1 strong line \& 1 weak line), 2=weak (1 strong line only or weak lines only), and 1=poor (1 weak line or continuum only).  For this example, Milliquas accepts only those objects with flags 3 or 4 as classified quasars.  The general rule is that one strong line plus something extra is required, that something extra being either a redshift-agreeing weak line or agreeing photometric redshift.  Many papers publish "quasars" with a mix of spectroscopic and photometric redshifts; in such cases, only the spectroscopic quasars are accepted into Milliquas as classified quasars because a confident redshift is required as well as a confident classification.  Thus those quasars that are published without redshift do not appear in Milliquas, unless qualifying as candidates.

\section{Milliquas since the publication of the Half Million Quasars catalogue in 2015}  

Milliquas has been published as a "live" catalogue updated at irregular intervals.  Two versions were published on arXiv due to significant changes:  Version 6.4 \citep{FL2019} presented an upgrade of its optical background data and other changes since the 2015 HMQ, and version 7.2 \citep{FL2021} provided a detailed accounting of the inclusion of the Sloan Digital Sky Survey (SDSS) Quasar Catalogue 16th Data Release \citep[DR16Q:][]{DR16Q} into Milliquas, in particular identifying DR16Q objects rejected from Milliquas and also those SDSS quasars left out of DR16Q but included into Milliquas.  Both papers also described the latest quasar discoveries to the time of publication.            
     
The NASA HEASARC Milliquas page gives a running summary of all the announced changes since version 0.1 in 2009.  All throughout, there was an unpublicized ongoing campaign of fixits and astrometric/photometric tightening which have honed Milliquas to its standard of accuracy today.  An early such exercise was to inspect legacy quasars which had non-quasar colours -- this yielded many cases of mistaken identity and the true quasar found and substituted.  The most recent large-scale exercise was to identify classified quasars which were flagged by \textit{Gaia}-EDR3 as having likely proper motion or parallax, and which had poor spectra (especially early grism spectra) and no radio/X-ray association, and so could easily be dropped.  About 110 net legacy quasars and 326 SDSS-DR16 \citep{DR16} pipeline-only quasars were dropped in the exercise, but 693 LAMOST\footnote{LAMOST QSO Catalog page: https://nadc.china-vo.org/data/article/20190107155838} objects, previously discounted as low-confidence, were restored as quasars due to \textit{Gaia}-EDR3 measuring them but finding no significant proper motion or parallax.  Thus the unexpected net outcome of the exercise was an $\approx$250 increase in the Milliquas quasar count.  Notable among the legacy deletions were 12 (out of 91) Cyril Hazard unpublished quasars (cite=0800), having neither radio/X-ray association nor redshift confirmation from SDSS, dropped due to proper motion detected by \textit{Gaia}-EDR3.   

Milliquas has always presented quasar candidates in addition to classified quasars.  The HMQ \citep{HMQ} candidates were only those of 99\% pQSO (confidence of being true quasars) calculated using photometry and required radio/X-ray association, and in this final edition Milliquas has returned to that selection.  In-between, Milliquas featured a much expanded repertoire of candidates.  Those radio/X-ray associated candidates now excluded from Milliquas are still available in the MORX v2 \citep[MORX:][]{MORX} catalogue.  Candidates without radio/X-ray association were dropped from Milliquas as of version 7.4 to avoid simply repeating candidates from other catalogues.

\section{Uptake of the DESI-EDR}

The Dark Energy Spectroscopic Instrument Early Data Release \citep[DESI-EDR:][]{DESEDR} is of data from their commissioning and Survey Validation prior to starting their DESI Main Survey.  Unfortunately their Quasar Redshift Value-Added Catalog (VAC) was not released in time for Milliquas v8, but I have processed the large (2\,847\,435 rows) DESI-EDR pipeline data along with three available VACs (\cite{DESVI} of 5496 rows, \cite{DESLyA} of 20\,281 rows, and \cite{DESBAL} of 83\,781 cleaned rows) to yield 91\,011 eligible DESI-classified quasars, of which, however, 1982 were rejected via my inspections of DESI spectra as described below, leaving 89\,029 accepted quasars of which 61\,342 are new quasars in Milliquas. 

The large DESI-EDR pipeline file presents a cleaned version of its total available pipeline data, the dropped objects being those bearing certain ZWARN (observation warning flag) values, including some I would have liked retained.  The 3 VACS use different criteria for data selection, so do have some data bearing those "forbidden" ZWARN values.  The VACs overlap somewhat, and shared objects can be classified differently, e.g., 467 objects are classified as galaxies by \cite{DESBAL} but classified as quasars by \cite{DESLyA}.  Similarly in the large pipeline file, 403 DESI-classified quasars were also DESI-classified as galaxies or stars, i.e., there were duplicate rows in the pipeline file for these targetIDs, with different classifications.  Inspection of some of these spectra revealed mostly stars or inscrutable profiles.  Rather than inspect all these dual-classified spectra, I dropped them all because my priority was to exclude false positives even if some true quasars were lost as well.  The presence of these dual classifications also brings to light that DESI's focus is not the absolute accuracy of each datum, but rather that bulk data (with an acceptable small error rate) will be used to test cosmological hypotheses.  However, Milliquas does aspire to the absolute accuracy of each datum, so visually inspecting DESI spectra and consequently editing redshifts \& classifications as appropriate, is necessary.  However, visually inspecting 91K DESI spectra is impractical, so a course of action is outlined below whereby those spectra are identified which are most likely to have pipeline redshifts discrepant from the true. 

DESI fits their spectral models onto 7 "morphtype" (visual morphology) classes, and each observed object is allocated to one of those morphtypes for spectral comparison against that model.  The 91\,011 eligible quasars defined above yield these counts per morphtype: 79\,559 objects with morphtype="PSF" (stellar point-spread function), 264 "GPS" (\textit{Gaia}-positioned PSF), 4939 "REX" (round exponential galaxy), 2292 "SER" (Sersic galaxy profile), 2287 "DEV" (DeVaucouleurs: elliptical galaxy), 524 "EXP" (exponential: spiral galaxy), and 4 "GGA" (\textit{Gaia}-positioned galaxy).  Also there are 1142 with blank morphtype, of which 549 are listed with DESI \textsl{grz} photometry, and 593 show no photometry.  I treat the blank morphtype with \& without photometry as two notional morphtypes in addition to the 7 DESI morphtypes, thus 9 morphtypes in all used in my method below for finding key spectra to visually inspect.                  

Notes on the processing of the DESI-EDR QSOs and the 1982 rejections:
\begin{itemize}
\item For each morphtype (see above) I ordered its objects by redshift to find redshift ranges which were not well-performed, i.e., showed unclear spectra.  When such ranges were identified, I did extensive visual checks on all spectra at the upper/lower bounds to keep or drop individual objects found there; in total, I did about 1000 visual DESI spectrum inspections.  
\item About 28K of the objects had been previously classified elsewhere.  I imported those classifications to label "truth" objects to identify reliable swathes of redshifts and less reliable ones.
\item Blank morphtype with photometry tended to fail for z<1.4, i.e., poor spectra.  Similarly, "PSF" morphtype with faint photometry (flux\_z<1) tended to fail for z<1.6.  I rejected most of those, bookending the rejected swathes with sequences of visually-inspected spectra which failed the inspection, plus random inspections within such swathes. 
\item Blank morphtype without photometry was excellent for z=3.082 to 4, i.e., all spectra clear and decisive, also for most of z=2 to 2.233.  Also its coverage of the COSMOS X-ray and Andromeda Commissioning fields was excellent.  Other redshift ranges were heavily inspected and some rejected.    
\item I visually checked all spectra of z$\ge$3.5 to be consistent with previous handling of SDSS spectra in Milliquas.  Those with unconvincing spectra were rejected. 
\item I visually checked or accepted all REX/EXP/DEV/SER/GGA morphtypes.  There are not many of these compared with the PSF/GPS morphtype for QSOs, but these AGN spectral models were well-performed although not at high redshifts (z$\ge$3.5).  
\item There are 119  M-star/white-dwarf doublets DESI-classified as quasars in the redshift range of 0.6433 to 0.6586, an example being targetID=39627992170762093\footnote{viewable at https://www.legacysurvey.org/viewer/desi-spectrum/edr/targetid39627992170762093.  To see any DESI spectrum, just replace the DESI target ID at the end of the foregoing web address.  Note that the spectrum is displayed in red, whereas the the black lines show model fit(s).}.  These are some crazy-looking spectra.  I dropped them all, of course.  
\item 50 DESI sources were just star-glow artefacts offset 1-2 arcsec from their true known centroid (thus a duplicate), an example being targetID=39628526864829382 (z=1.4106) which does not exist and is offset 1.273 arcsec from quasar SDSS J094234.47+322035.6, z=1.409.  These 50 artefacts were dropped. 
\item I was reluctant to accept spectra which had only negative or inverted lines.  They looked mostly random to me, and not like BAL-type lines.  So I may have missed valid quasars of that type.    
\item I have assigned new redshifts to 3 DESI QSOs: 
\begin{itemize}
   \item 39633247998578429, DESI z=0.6528, changed to z=1.94, obvious Ly$\alpha$-SiIV-CIV-CIII lines, dual qso-star.  
   \item 39633161977597561, DESI z=1.6632, changed to z=0.747, 7 lines matched, especially the Mg-II \& O-II duo. 
   \item 39632930619786218, DESI z=1.1321, changed to z=2.79, obvious Lyman forest, emission lines match ok, anomalous absorption lines throughout the spectrum are most mysterious.  
\end{itemize}
\item I selected DESI-classified galaxies and stars which MORX shows to be radio-associated, which is a good predictor of AGN activity.  Of those, 41 DESI galaxies and 6 DESI stars have been assigned new redshifts and "promoted" to quasars, upon my inspection of those spectra, an example being 39628514877505752, DESI galaxy of z=1.0408, changed to z=2.376 with obvious Ly$\alpha$ and other lines.  Also there is 39628433285714226, a DESI star, changed to z=2.235 with 7 lines matched, although you need to move the "Gaussian Sigma Smooth" slider over to see them -- showing how well the smoother reveals faint lines otherwise concealed by noise.  These 47 new quasars can be identified in the Milliquas data by cite='DESEDR' and zcite='MQ'.  
\item A further 101 DESI galaxies have been "promoted" to AGN, and 53 DESI galaxies "promoted" to narrow-line (type-II) AGN by virtue of radio/X-ray association and active spectral lines, especially when both Mg-II and O-II are prominent.  The DESI redshift is unchanged for these.  An example is 39633332224397332.  These can be identified in the Milliquas data by cite='DESEDR' and zcite='DESEDR' and classification of A or N, and the comment includes an 'a' (host-dominated).      
\item 5 line poachers\footnote{"line poachers" are introduced with multiple examples in \citet{FLES40}, Section 5.}  were found, these are stars with phantom spectral lines acquired via blended glare from a nearby quasar or galaxy; they are easily recognized because they have very different optical colours compared to the neighbouring quasar with which they putatively share the same redshift.  They are: 
\begin{itemize}
\item 39632955013858840 with putative z=0.5763 at 1.295 arcsec offset from quasar SDSS J075807.01+334955.9, z=0.576. 
\item 39628281896501832 with putative z=0.4250 at 2.089 arcsec offset from quasar 3C 47.0, z=0.425.       
\item 39628373638516357 with putative z=1.6683 at 1.108 arcsec offset from quasar SDSS J125703.80+250457.5, z=0.821, which is also the correct redshift for the DESI spectrum once the Mg-II line is moved to the obvious bump at 5100\AA\ which also aligns the O-II and other lines. 
\item 39633132038653380 with putative z=0.3977 at 1.075 arcsec offset from galaxy SDSS J154931.53+425905.0, z=0.398.  This object looks like a K star.     
\item 39633352331887198 with putative z=1.8645 at 5.291 arcsec offset from quasar SDSS J120930.54+572050.6, z=1.867.  Whilst this is quite far offset for a line poacher, the DESI object spectrum shows no feature apart from the lines faintly echoing the SDSS quasar; possibly it is a galaxy faintly lensing the quasar.
\end{itemize} 
These 5 line poachers were dropped from Milliquas processing.     
\end{itemize}
As a final note on DESI pipeline redshifts, they are very good but sometimes they wrongly match many minor lines while missing the "elephant in the room", i.e., the Mg-II double-line which often presents as a large thick mound in the spectral profile.  Confirming its identity is the O-II double-line at a standard offset from the Mg-II lines -- the Mg-II wavelength is at 75.1\% of the O-II wavelength, and the O-II lines are typically used to refine the redshift precisely.  Thus this Mg-II/O-II combo can signal a precise mid-range redshift.  An example of the informative power of this combo is 39633413539368493 which is an AGN but with a spectrum showing a strong galaxy continuum which conceals the AGN lines\footnote{to see this clearly, go to https://www.legacysurvey.org/viewer/desi-spectrum/edr/targetid39633413539368493, set the Gaussian Sigma Smooth slider to 4, and turn off the black pipeline model by pressing the "pipeline fit" and "other model" labels.  The spectrum is now seen as basically featureless.} for which the DESI pipeline has settled on z=1.6366.  Inspection shows a possible swelling at 6100\AA\ as a candidate Mg-II line, so the redshift sliders can be used to bring the Mg-II lines over to it.  Having done so, the O-II line is seen to align pleasingly to a small spike in the spectrum, so tweaking the redshift slider exactly there reveals a redshift of 1.169.  Confirmation comes at the left of the spectrum where the C-III line now aligns to a subtle spectral mound.          

The upcoming DESI Quasar Redshift VAC will provide an "Mg-II afterburner" redshift in addition to the pipeline redshift, using the broad Mg-II line as the anchor; presumably the O-II line will also be used to refine the redshift.  Also DESI will issue a QuasarNET VAC based on the  hierarchical-heuristic redshift-finding algorithm of \cite{QuasarNET}, and use all these to decide the quasar redshifts for their cosmological investigations.  In lieu of those future tools, I have used the above method of targetted visual inspections to decide the redshifts of these DESI spectra.

\section{Uptake of the SDSS-DR18Q}          

The Sloan Digital Sky Surveys DR18 \citep[DR18:][]{DR18} was issued in January 2023 and came with pipeline data and its "Black Hole Mapper" visual VAC \citep[DR18Q:][]{DR18Q} data for which, however, the VAC paper has not yet been made available as of July 2023.  The DR18 approach has changed from earlier releases in that presented data is of newly observed objects only, and the visual VAC covers all of the pipeline quasars.  The DR18Q consists of 3 files, one being the main file of visually inspected spectroscopic redshifts and classifications, and the other two appear to be working redshift data of uncertain status, with their paper unavailable to clarify.  Thus I used only the main visual file.   

The file is of 13085 rows of which 6408 are classifed as quasars, all of which are in the pipeline data also.  Processing these yielded 4551 new type-I quasars/AGN, 93 new type-II (because subclassed as STARRY and not BROADLINE), and 1763 objects previously catalogued in Milliquas.  There is also one new line poacher: SDSS J083217.11+040403.9 with putative z=1.123 at 2.257 arcsec offset from quasar SDSS J083216.99+040405.2, z=1.121.  This line poacher was dropped from Milliquas processing\footnote{For the record, there was also a new line poacher in SDSS-DR17 \citep{DR17}: SDSS J084856.10+011538.7 with putative z=0.645 at 1.264 arcsec offset from quasar SDSS J084856.08+011540.0, z=0.646.}.   
   
There were 2 pipeline quasars not included in the visual file; why the omission, is unclear.  Those 2 objects have quasar-like photometry and pipeline redshifts, and I have included them into Milliquas trusting that they are valid quasars and that their absence from the visual file was inadvertent.  They can be identified in the Milliquas data by cite='DR18' and type='Q'.

\begin{table*} 
\scriptsize	 
\caption{Five new quasars presented in this paper}
%\fontsize{4.5pt}{5.5pt}\selectfont
\tiny
%\begin{tabular}{rlclcrrl}
\begin{tabular}{@{\hspace{0pt}}r@{\hspace{4pt}}l@{\hspace{6pt}}c@{\hspace{4pt}}l@{\hspace{6pt}}c@{\hspace{4pt}}r@{\hspace{4pt}}r@{\hspace{4pt}}l}
\hline 
\# & new quasar ID & new quasar \it{ugriz} & prev ID (doublet partner) & previous ID \it{ugriz} & offset & redshift & comment \\
\hline
 1 & SDSS J002620.66-034135.8 & 21.95 21.89 21.44 21.34 21.46 & SDSS J002620.72-034134.2 & 24.62 22.09 20.77 19.86 19.38 & 1.88" & 1.605 & bluish SDSS stellar PSF, prev \\ 
   & & & & & & & ID is reddish SDSS "galaxy" \\ 
 2 & SDSS J080212.45+285927.5 & 22.10 22.17 21.56 21.67 21.55 & SDSS J080212.35+285928.3 & 23.48 22.56 21.46 20.31 19.99 & 1.52" & 1.430 & radio ILT J080212.41+285927.0 \\ 
 3 & SDSS J105030.49+200420.9 & 20.26 19.57 20.16 19.41 19.34 & SDSS J105030.51+200422.1 & 21.13 20.64 18.76 18.31 18.00 & 1.23" & 1.650 & SDSS-DR7 photometry, \textit{Gaia} \\  
   & & & & & & & 3$\sigma$ proper motion for prev ID \\ 
 4 & SDSS J134915.81+080120.3 & 19.60 19.49 19.36 19.19 18.83 & SDSS J134915.76+080122.5 & 24.10 22.35 20.46 19.56 19.14 & 2.41" & 2.038 & X-ray 4XMM J134915.8+080121  \\ 
 5 & SDSS J151135.04+495958.2 & 24.65 21.51 21.49 21.67 21.48 & SDSS J151135.19+495958.1 & 23.29 22.59 20.55 19.86 19.48 & 1.60" & 2.709 & radio ILD J151135.08+495958.0  \\ 
   & & & & & & & (ILT Gaussian), \textit{Gaia}-EDR3 \\   
   & & & & & & & 20$\sigma$ proper motion for prev ID \\ 
\hline 
   & & & & & & &  \\  
\end{tabular}
\end{table*}

\section{Five new quasars presented by this edition}

In \cite{FLES40} I presented 40 new quasars discovered by searching the total SDSS-DR16 data, 28 of which had been confused with or concealed by stars in close doublets on the sky.  Here I present 5 more such new quasars revealed within close star-quasar doublets for which SDSS identified the star as being the quasar.  Table 2 lists the 5 new quasars and their doublet partners, showing the \textit{ugriz} photometry for each.  In each case, the new quasar shows relatively flat photometry (with the bottom quasar also showing \textit{u}-dropout consistent with its redshift) whilst the previous ID shows colours typical of a red star.  The "comment" column shows radio/X-ray detections for 3 of the doublets, each positionally better aligned to the new quasar than the previous ID; also 2 of the previous IDs are assigned proper motion by \textit{Gaia}-EDR3.  The 3rd doublet photometry is sourced from SDSS-DR7 \citep{DR7}, because SDSS-DR16 does not resolve this doublet.  The 5 SDSS identifications were basically "line poachers" with the true quasars not identified; but now, this paper identifies the true quasars.   

Figure 2 shows images of the 5 new quasars, with an explanatory caption.  The example spectrum shows typical features of such a combined spectrum; spectra of the 4 other doublets are shown in \cite{FL2021} Figures 11 \& 12 with explanatory captions.  In each case SDSS-DR16 processing had confused the star with the quasar, and the current SDSS-DR18 has not updated the identification.

\begin{figure} 
\includegraphics[scale=0.25, angle=0]{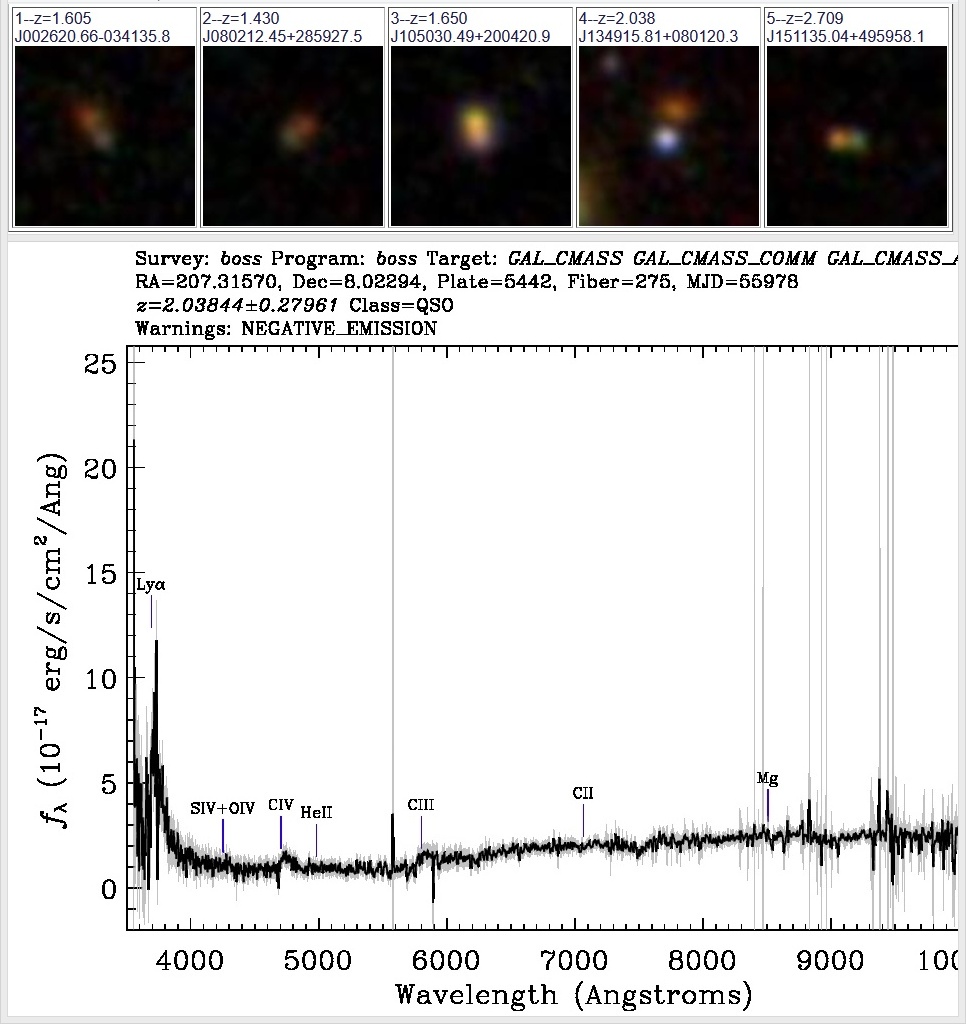} 
\caption{The 5 new quasars presented in this paper, on SDSS images with 15 arcsec sides.  Each new quasar is at the exact centre of its image, with its doublet partner (which was classified by SDSS as the quasar) alongside.  For each doublet, the merged spectrum shows quasar lines at the blue end where the quasar's flux dominates the star's flux, and red stellar continuum dominating the right side.  The joint SDSS spectrum of the 4th doublet (displayed, z=2.038) reveals the Ly$\alpha$ line dominating the blue end at left. \\} 
\end{figure}

\section{QSOs dropped since Milliquas v7.2}

A recent prominent exercise to drop previously-included quasars from Milliquas was the test against \textit{Gaia}-EDR3 movers described in Section 3.  However, that test was not decisive on its own because there are always anomalies in large data, so patterns need to be checked for.  In this case some of these apparent movers were found to be star-quasar superpositions, or a nearby moving star which caused the observed optical PSF centroid of the quasar to subtly shift within the changing star flux gradient, or that same outcome can happen if either member of a close star-quasar doublet is optically variable.  In such cases the quasar can be valid even though flagged by \textit{Gaia}-EDR3 as a mover.  Thus each object so flagged was inspected prior to deletion -- if it had neither a convincing spectrum nor radio/X-ray association, nor was in a close doublet, it was dropped.  Close doublets warranted additional checks to confirm a valid quasar.   

Also, Milliquas excludes low-confidence/quality or questionable objects (so deemed by their researchers), but many such objects were inherited from VCV \citep{VCV} which was more forgiving.  They were removed as encountered (after being checked over as described in the preceding paragraph) but a residue remains.  Sometimes such objects would be found to be duplicates of nearby quasars, or positioned onto a faint galaxy with no quasar-like object nearby nor at due N/S or E/W offsets (such offsets occasionally found in 20$^{th}$ century publications).  Table 3 gives 14 such individual deletions of legacy "quasars" since Milliquas v7.2 \citep{FL2021}, with explanations.

\begin{table*}
\scriptsize	 
\caption{14 deletions of legacy quasars since Milliquas v7.2}
\tiny
\begin{tabular}{rlllll@{\hspace{6pt}}l} 
\hline 
\# & Name & J2000 & rmag & z & OA ref$^{*}$ & comment on deletion \\
\hline
1 & PHL 1222 & 01 53 53.89 +05 02 57.1 & 17.57 & 1.904 & 0282 & a star, conflated with faint quasar SDSS J015354.03+050259.7 unseen by OA$^{\dagger}$. \\
2 & 1WGA J2204.9-1815 & 22 04 51.91 -18 15 36.4 & 21.08 & 0.210 & 0307 & duplicate of blazar 1REX J220451-1815.5 at 8.6 arcsec offset. \\
3 & NGP9 F324-1105786 & 13 38 43.10 +26 25 02.6 & 18.03 & 1.260 & 0424 & image shows galaxy, OA said "uncertain" redshift. \\
4 & Q J1943.9-1502 & 19 43 58.84 -15 02 48.0 & 18.79 & 3.300 & 0429 & fuzzy, \textit{Gaia}-EDR3 proper motion. \\
5 & Q 0057-2811 & 01 00 23.80 -27 55 46.1 & 19.61 & 2.120 & 0530 & image shows galaxy, redshift wrong for it. \\
6 & Q J08500+701 & 08 50 02.32 +70 18 06.6 & 14.08 & 1.900 & 0994 & obvious member of star group, X-ray is from NGC 2650. \\
7 & XFLS J17178+5847 & 17 17 50.73 +58 47 44.0 & 23.20 & 2.550 & 1041 & OA stated '?' for classification and redshift. \\
8 & SW002802.79-425957.0 & 00 28 02.79 -42 59 57.0 & 21.25 & 1.731 & 1042 & red object, OA stated obscured and that class \& z were not secure. \\
9 & Abell 2690\#075 & 23 59 56.60 -25 10 20.0 & 25.00 & 2.130 & 1145 & OA stated "tentative" classification \& redshift. \\
10 & QNY1:29 & 12 36 27.94 -00 53 57.5 & 20.05 & 0.179 & 1534 & classified as galaxy by SDSS-DR17. \\
11 & 16V13 & 08 44 50.55 +44 35 53.9 & 22.53 & 1.833 & 1551 & absorption spectrum only, QSO "WEE 13" is 18 arcsec NW, possible confusion. \\
12 & 1RXS J104057.7+134216 & 10 40 57.68 +13 42 11.6 & 22.18 & 0.556 & 4LAC & duplicate of QSO SDSS J104058.37+134150.6, z=0.557, at 23 arcsec SE. \\
13 & LEDA 138501 & 02 09 34.59 +52 26 32.6 & 18.54 & 0.049 & BASS & duplicate of QSO 1ES 0206+522, z=0.049, at 26 arcsec NE. \\
14 & CGCG 70-200 & 12 38 22.39 +09 31 41.9 & 13.82 & 0.321 & PGC & SDSS finds non-AGN. \\
\hline
\multicolumn{7}{l}{$^{*}$ references are indexed in the accompanying file "milliquas-references.txt".} \\
\multicolumn{7}{l}{$^{\dagger}$ OA = original author(s) of discovery paper.} \\ 
   & & & & & &  \\
   & & & & & &  \\  
\end{tabular}
\end{table*}

\section{Radio/X-ray associations presented}   

Milliquas has always presented radio/X-ray associations for its objects.  The likelihoods for these associations are as calculated by a data-driven algorithm which compares areal densities of photometry-association combinations with backgound averages; this is documented in detail in my previous papers, especially \cite{QORG} which you would read at your peril.  The likelihoods of individual radio/X-ray associations are no longer displayed in Milliquas (as the final Milliquas projects pQSO$\ge$99\% for all its objects) but are displayed in the MORX \citep{MORX} catalogue which itemizes radio/X-ray associations for > 3 million optical objects over the whole sky.  All Milliquas v8 objects having radio/X-ray associations are reported in MORX also; both final catalogues were extracted out of the same underlying large database which was frozen after processing all data to 30 June 2023.  

The MORX paper comprehensively describes the large radio and X-ray surveys used there and in Milliquas.  Here I list just a brief overview of changes since Milliquas v7.2 \citep{FL2021}: 
  
\begin{itemize}
  \item Radio/X-ray association likelihoods are now calculated at a granularity of 0.1 arcsecond astrometric offsets for all radio/X-ray source catalogs.  This wouldn't be necessary for low-resolution early catalogues such as those from \textit{ROSAT} but was done so that uniform onwards processing could be used for all.    
  \item VLASS \citep{VLASS} has been reprocessed to include only Gaussian detections, and its inclusion threshold lowered from S/N=5 to S/N=4.  The effects are better and more detections, respectively.   
  \item RACS \citep{RACS} radio associations have been added.
  \item LoTSS \citep{LoTSS} radio associations have been added.  This was a very large input catalogue.  
  \item RASS (\textit{ROSAT} All-Sky Survey) X-ray data has been dropped, as it has become clear over time that its resolution is too coarse to confidently identify optical sources in isolation.
\end{itemize}

\section{Overview of Milliquas -- data and structure}

Table 4 shows the Milliquas data structure.  Each object is displayed with J2000 astrometry (usually from \textit{Gaia}-EDR3 precessed to J2000 by CDS\footnote{https://cds.u-strasbg.fr/}), red and blue photometry, redshift, citations, and radio and X-ray associations where present.  The intent is to present simple data of one line per reliable object.  The ReadMe elaborates on the structure and indexes the values found in each field.      

\begin{table*} 
\caption{Sample lines from Milliquas} 
\includegraphics[scale=0.26666, angle=0]{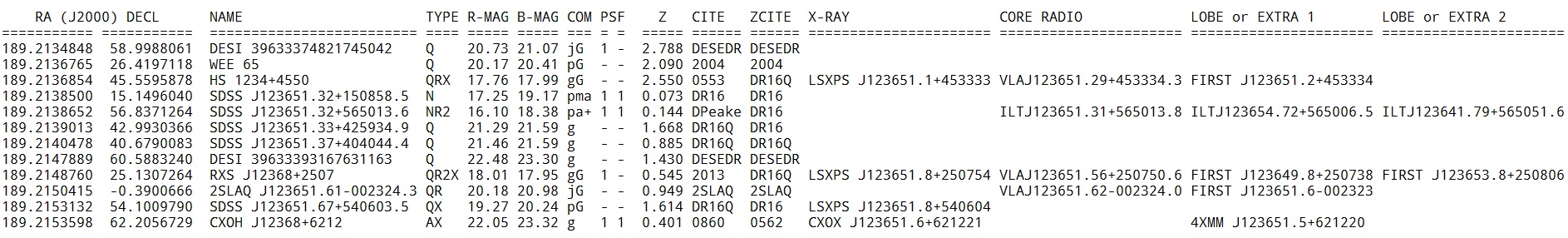} 
\tiny \
Notes on columns (see ReadMe for full descriptions): 
\begin{itemize}
  \item TYPE: 1st char is the object classification (if classified): Q=QSO, A=AGN, N=type II AGN, see ReadMe for full list.  Chars summarizing the associations displayed: R=radio, X=X-ray, 2=double radio lobes. 
  \item COM:  comment on photometry: p=POSS-I magnitudes, so blue is POSS-I \textsl{O}, j=blue is SERC \textsl{Bj}, g=SDSS \textsl{g} \& \textsl{r}, +=optically variable, G=\textit{Gaia}-EDR3 astrometry, a=faint nuclear activity, m=nominal proper motion.
  \item PSF:  for red \& blue sources: '-'=stellar, 1=fuzzy, n=no psf available, x=not seen in this band.
  \item CITE \& ZCITE: citations for name and redshift; citations are indexed in "milliquas-references.txt".
  \item LOBE or EXTRA: if TYPE contains a '2' (=lobes), then double radio lobe identifiers are displayed here.  Otherwise, any additional radio and/or X-ray identifiers are displayed here.
\end{itemize}
The full table can be downloaded from http://quasars.org/milliquas.htm. 
\end{table*}

Table 5 shows the top 25 contributing discovery papers ordered by numbers of name citations.  SDSS-DR16 catalogues dominate with 74\% of all quasar discoveries, but the large candidates catalogues also show prominently. 

\begin{table*} 
\centering
\scriptsize	 
\caption{Top 25 discovery papers for Milliquas v8}
\begin{tabular}{rlrrrl}
\hline 
\  &    &  \# of  & \# of & \# of & \\
\# & ID & classified & candidates & redshifts & paper \\
\hline
 1 & SDSS DR16Q visual       & 716639 &   265 & 394077 & \cite{DR16Q}   \\
 2 & DESI EDR                &  60095 &     1 &  64603 & \cite{DESEDR}  \\
 3 & SDSS DR16 pipeline      &  40898 &    96 & 373997 & \cite{DR16}    \\
 4 & 2QZ/6QZ                 &  27490 &    14 &  23260 & \cite{CROOM04} \\
 5 & PGC$^{\dagger}$         &  20347 &       &      8 & \cite{PGC}     \\
 6 & NBCKDE-v3 candidates    &    231 & 17474 &  15257 & \cite{NBCKv3}  \\
 7 & GAIA3 candidates        &        & 16526 &  18962 & \cite{GQC}     \\
 8 & XDQSO candidates        &        & 13963 &        & \cite{XDQSO}   \\
 9 & Milliquas v8            &      5 & 11349 &  20750 & data unique to Milliquas \\
10 & 2SLAQ                   &  10362 &     4 &   8259 & \cite{2SLAQ}   \\ 
11 & LAMOST QSO DR5          &   8131 &       &   7933 & \cite{LAM5Q}   \\
12 & LAMOST QSO DR3          &   7181 &       &   6846 & \cite{LAM3Q}   \\
13 & LAMOST QSO DR9          &   4875 &       &   4780 & \cite{LAM9Q}   \\
14 & SDSS DR18Q visual       &   4731 &       &   4731 & \cite{DR18Q}   \\
15 & AllWISE candidates      &        &  3558 &        & \cite{SECREST} \\
16 & NBCKDE candidates       &      1 &  2208 &   1990 & \cite{NBCKDE}  \\
17 & SDSS DR7Q visual        &   1969 &       &    200 & \cite{DR7Q}    \\
18 & AGES survey             &   1849 &     2 &   1023 & \cite{AGES}    \\
19 & 6dF                     &   1756 &       &    234 & \cite{6dF}     \\
20 & DEEP2 Redshifts         &   1544 &       &   1481 & \cite{DEEP2}   \\
21 & SDSS DR14Q visual       &   1528 &       &   1530 & \cite{DR14Q}   \\ 
22 & AAOz (XXL-South)        &   1491 &       &   1498 & \cite{AAOz}    \\
23 & OzDES2                  &   1218 &       &   1239 & \cite{OzDES2}  \\
24 & HETDEX                  &   1144 &       &   1009 & \cite{HETDEX}  \\
25 & DESI VI VAC             &   1117 &       &   1391 & \cite{DESVI}    \\
\hline
\multicolumn{6}{l}{Candidates from papers of >1000 classified objects, are all radio/X-ray associated objects which either have an} \\ 
   & \multicolumn{5}{l}{insecure classification / redshift (ZWARNING=4 in the case of the SDSS objects) or are classified stars} \\ 
   & \multicolumn{5}{l}{with quasar-like photometry and so retained as QSO candidates.} \\
\multicolumn{6}{l}{$^{\dagger}$ The PGC (Principal Galaxy Catalogue, a.k.a. LEDA) is not actually a discovery paper, but is used to source} \\
   & \multicolumn{5}{l}{names for AGN galaxies.} \\
\end{tabular}
\end{table*}

\section{Conclusion} 

The Milliquas (Million Quasars) catalogue v8 is presented as a complete record of published quasars to 30 June 2023.  Milliquas presents 907\,144 type 1 QSOs \& AGN, 66\,026 high-confidence (pQSO=99\%) radio/X-ray associated quasar candidates, 2814 BL Lac objects, and 45\,816 type 2 objects.  Astrometry is 0.01 arcsecond accurate for most objects, and red-blue photometry is of 0.01 magnitude precision.  Radio and X-ray associations for these objects are presented as found, including double radio lobes.  This is the final edition of Milliquas.

In future, when quasar counts have gone over 2 million and Milliquas is increasingly just a relic of the past, its enduring deliverable will be its accurate positioning of all the legacy quasars.  Any future cataloguing effort should incorporate that.

\section{A personal note}

Born in the Netherlands, I made my way to New Zealand in 1984 with an engineering degree with a physics minor from the University of Colorado.  My employment started in engineering, but settled into computer programming and databasing.  When I got onto the internet in 1995, I sought out physicists and astronomers on Usenet, astronomy being my earliest interest.  I learned a lot from those discussions: facts, current thinking, and how to better present my ideas.  In those days I was interested in the cosmology of Halton Arp, and needing data to test that, I waited impatiently for someone to combine the optical, radio, and X-ray sky catalogues of the day.  But this was not forthcoming, and I realized that astronomers had their own specialties to pursue, and combining catalogues was not one of them.  So I undertook to do that myself.  I remember well my first download of an APM \citep{APM} optical file (which took an hour, and there were 958 of them), it was my "alea jacta est" moment.  I am grateful to Dave Monet for sending me CD's of his USNO-A1.0 and USNO-A2.0 optical catalogues which enabled me to process over the whole sky.  In the end, the new data did not support Arp's interpretation of quasar origins, so I abandoned that, but meanwhile was well in-gear with my all-sky processing.  I gradually added quasar cataloguing to my repertoire, starting in Y2000 when Mira V\'eron-Cetty invited me to help check over the pending VCV 9th edition data, which I did through to their final (13$^{th}$) edition.  I have since published papers and catalogues, and can attest that two key requirements for successful astronomy publication are a mathematically-grounded science education, and perseverence.

\section*{Acknowledgements}
Thanks to my tolerant wife.   This work was not funded.

\section*{Data Availability}
Milliquas can be downloaded from CDS at https://cdsarc.cds.unistra.fr/viz-bin/cat/VII/294 or from its home page at https://quasars.org/milliquas.htm which also provides a FITS file.  Both sites also provide the ReadMe and the references list.  Query pages are provided by CDS and NASA HEASARC at https://heasarc.gsfc.nasa.gov/W3Browse/all/milliquas.html.

%%%%%%%%%%%%%%%%%%%% REFERENCES %%%%%%%%%%%%%%%%%%

\label{lastpage}
\end{document}